\documentclass[jkps,fleqn,showpacs,showkeys]{revtex4}
\usepackage[pdftex]{graphicx}
\usepackage{amssymb}
\usepackage{amsmath}
\usepackage{bm}
\usepackage{soul}

\begin{document}
\setcounter{page}{1}
\title[]{Laser cooling of molecules}
\author{Eunmi \surname{Chae}}
\email{echae@korea.ac.kr}
\affiliation{Department of Physics, College of Science, Korea University}

\begin{abstract}

A recent progress on laser cooling of molecules is summarized. 
Since the development during 1980s for atomic species, laser cooling has been the very beginning step to cool and trap atoms for frontier research on quantum simulations, quantum sensing and precision measurements. 
Despite the complex internal structures of molecules, laser cooling of molecules have been realized with the deepened understanding of molecular structures and interaction between light and molecules. 
The development of laser technology over the last decades have also been a great aid for laser cooling of molecules because many lasers are necessary to successfully cool the molecules. 
A detailed principle and development of laser cooling of molecules as well as the current status of the field are reviewed to give an introductory of the growing field of ultracold molecular physics. 

\end{abstract}


\keywords{Molecules, Laser cooling, Magneto-optical trap, Sub-Doppler cooling}

\maketitle

\section{INTRODUCTION}
Laser cooling is a technique that cools down particles using momentum kicks of photons. 
Since its development using alkali atoms in 1980s, laser cooling opened a whole new era in atomic physics, ranging from quantum simulations to precision measurements.
Acknowledged the significance, three pioneers who developed the scheme received Nobel prize in physics in 1997.

The key for laser cooling is spontaneous emission from which energy dissipation of atoms originates. 
Therefore, laser cooling requires only one simple property to its objective - having a strong and closed cycling transition so that photon scattering events can be repeated many times. 
Alkali atoms with simple energy structures were excellent targets for laser cooling. 
With deepened understanding of the atomic structures and development of lasers, atoms with more complex energy structures such as alkaline-earth, Lanthanide, and group III atoms\cite{Yu2022} have also been laser-cooled and trapped in magneto-optical traps.

Success in laser cooling of atoms naturally has lead many interests to its extension to neutral molecules.
Molecules' large electric dipole moments and abundant internal structures make them an attractive platform for quantum simulation/computation, quantum chemistry, and precision measurements\cite{Carr2009, Bohn2017}.
Ultracold temperature is a prerequisite for most of these exciting research and laser cooling is one way to reach ultracold temperatures.

However, the simple requirement for laser cooling -- having a strong closed-cycling transition -- is unattainable for molecules due to their complex internal structures.
Molecules consist of multiple atoms so that they have additional degrees of freedom compared to atoms. 
They are vibration (quantum number $v$)  and rotation (quantum number $R$) of the constituting nuclei. 
Transitions between rotational states obey selection rules, and one can form a closed cycling transition by exciting the molecules from states with rotational quantum number $R=1$ to states with $R=0$.  
However, there are no selection rules for vibrational states. 
A transition probability from one vibrational state to another is only determined by the overlap of the involved vibrational wavefunctions, whose absolute square is called Franck-Condon factors (FCFs). 
When a molecule in a particular vibrational state in the electronic ground state is excited by a laser to a state in the electronic excited state, the molecule can decay into many vibrational states in the electronic ground state depending on its FCFs. 
Once the molecule decays into states other than its initial state, the molecule cannot absorb the laser anymore and the optical cycling ceases.
To overcome this problem, one would need to add many lasers to put the molecule back to the optical cycling.
Therefore, it has been considered for a long time that laser cooling of molecules is not feasible.

The flow changed when people found that some diatomic molecules have highly diagonal FCFs\cite{Rosa2004, McCarron2018a, Fitch2021a}. 
When excited, these molecules predominately decay to their initial states. 
By adding a few lasers to repump some "leakage", one can form a quasi-cycling transition that can achieve enough photon scattering to cool the molecules down to the Doppler limit temperatures. 
Based on this idea, one dimensional laser cooling, longitudinal laser slowing, and eventually a magneto-optical trap (MOT) for molecules have been demonstrated successfully.
Also, this idea can be extended to polyatomic molecules as well, and a MOT for CaOH molecules has been achieved recently\cite{Vilas2022}.
Table \ref{table:MoleculeList} summarizes molecules that have been laser-cooled or are on-going to be laser-cooled.

\begin{table}
\caption{On-going experiments on laser cooling of neutral molecules based on publications}
\begin{ruledtabular}
\begin{tabular}{ccc}
 Molecule & Current status & References \\
 \hline
 SrF & MOT achieved & \cite{Barry2014, Norrgard2016} \\
 CaF & MOT achieved & \cite{Anderegg2017, Truppe2017} \\
 YO  & MOT achieved & \cite{Collopy2018} \\
 YbF  & 1D cooling achieved & \cite{Lim2018} \\
 BaH  & 1D cooling achieved & \cite{McNally2020} \\
 CaH  & 1D cooling achieved & \cite{Vazquez-Carson2022} \\
 BaF  & 1D cooling achieved & \cite{Zhang2022} \\
 AlF  & Deflection achieved & \cite{Hofsass2021} \\
 MgF  & Deflection achieved & \cite{Gu2022} \\
 AlCl  & Spectroscopy done & \cite{McCarron2021, Daniel2021} \\
 TiF  & Spectroscopy done & \cite{Norrgard2017} \\
 CH & Spectroscopy done& \cite{Schnaubelt2021}\\
 CaOH  & MOT achieved & \cite{Vilas2022} \\
 YbOH  & 1D cooling achieved & \cite{Augenbraun2020} \\
 SrOH  & Deflection achieved & \cite{Kozyryev2017} \\
 CaOCH$_3$  & 1D cooling achieved & \cite{Mitra2020} \\
 SrOC$_6$H$_5$ & Spectroscopy done & \cite{Lao2022}
\label{table:MoleculeList}
\end{tabular}
\end{ruledtabular}
\end{table}

After cooled down to below 1 mK,  thermal excitations can be neglected and molecules are populated in a few quantum states.
This means molecules can now be treated just as atoms.
Sub-Doppler cooling is applicable to molecules and molecules can be transferred to a conventional trap such as a magnetic trap or an optical dipole trap.
Their quantum states can be manipulated using lasers, microwaves, and DC electromagnetic fields. 
With the success of laser cooling, people now are pursuing interesting physics with these molecules -- realizing long coherence time in optical tweezers\cite{Burchesky2021}, controlling collisions using molecular quantum states and external fields\cite{Anderegg2021}, and making quantum gate operations\cite{Holland2022}.
In this review, we summarize the development of laser cooling of diatomic molecules, molecules with the simplest structures.  
At the last part, current experiments with laser-cooled molecules and extensions to polyatomic molecules are briefly discussed as well. 


\section{Forming a quasi-cycling transition of molecules and deflection of a molecular beam}

\begin{figure}[t!]
\includegraphics[width=15.0cm]{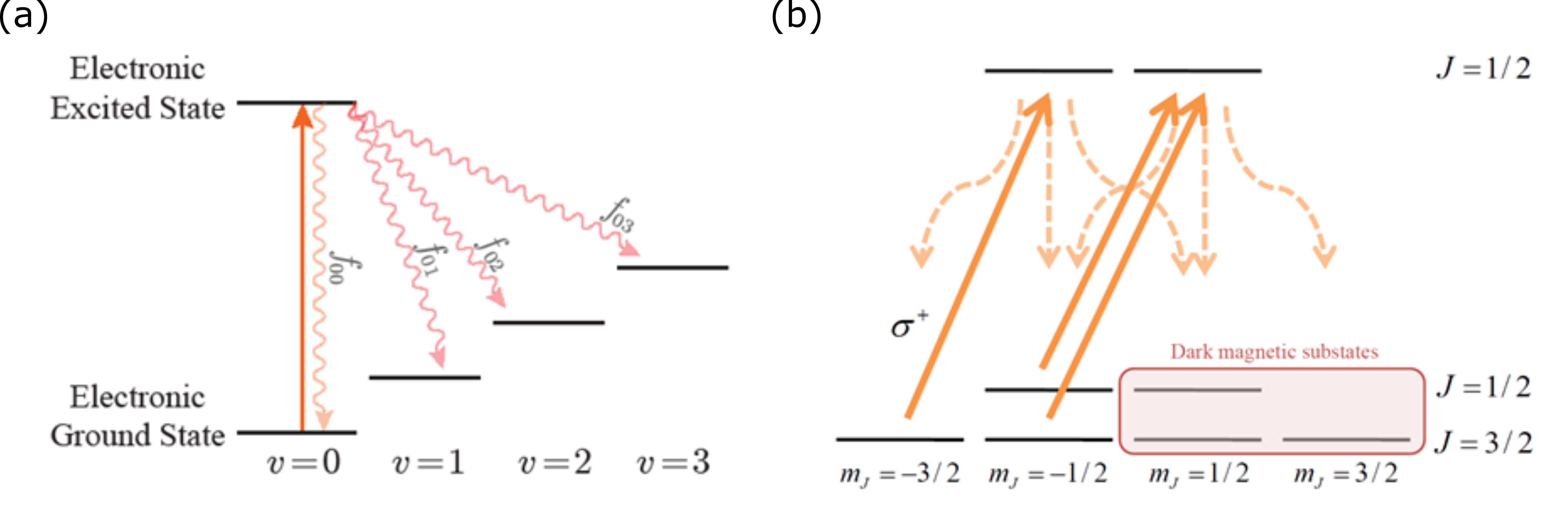}
\caption{Challenges for laser cooling of molecules. (a) Non-diagonal vibrational decays (b) Dark magnetic substates} 
\label{Fig:Challenge}
\end{figure}

The degrees of freedom related to the relative motions of nuclei in molecules - vibrations and rotations - hinder the molecules from forming closed cycling transitions.
A transition probability from an electronic excited state with a vibrational quantum number $v'$ to an electronic ground state with $v$ is proportional to a corresponding FCF, a square of the vibrational wavefunction overlap of the two states. 
Because the energies of the involved two electronic states depend differently on the distance between the two nuclei of the diatomic molecule, the overlap of the vibrational wavefunctions with different quantum number $v$s are usually non-zero, leading to non-zero probabilities for the excited molecule to decay into states with different $v$s (Figure \ref{Fig:Challenge}(a)). 
This inherently forbid the optical cycling, which requires the particle to always decay back to its initial state after being excited.  

However, there exist number of molecules whose FCFs are relatively diagonal, which means that the electronically excited molecules with $v'$ decay back to the electronic ground state with the same vibrational quantum number $v=v',$ with high probabilities of about 95\% -- 99\%\cite{Rosa2004}.   
These molecules can scatter about $10^5$ photons when additional two or more lasers are added to repump the vibrational leakage back to the optical cycling.
The number of photon-scattering achieved in this "quasi-cycling" transitions is enough to cool the molecules down to Doppler limit and trap them in a MOT.
Authors in Ref.\cite{Rosa2004} listed diatomic molecules that are predicted to have diagonal FCFs.
Recent studies have revealed that even some big molecules with optical cycling center are expected to have good FCFs and thus suitable for laser cooling\cite{Augenbraun2020a, Lao2022, Yu2022, Dickerson2022}.

Dealing with the rotational structure of molecules is simpler. 
Transitions between the rotational states obey selection rules just as other angular momenta.  
One caution is that the rotational quantum number $R$ starts from zero and expands to infinity for all states.
When a molecule is in the electronic ground state with $R=0$ initially, the electronically excited state should have $R'=1$ after the molecules absorbed a photon.
According to the selection rules, this excited state can decay back to the electronic ground state with either $R=0$ or $R=2$, therefore forming a non-closed cycling transition. 
For the molecules to continue scattering photons, one should employ a transition from an electronic ground state with $R=1$ to an electronic excited state with $R=0$\cite{Stuhl2008}.

In this configuration, there are more number of substates in the ground state than in the excited state, resulting in dark magnetic substates when the molecule scatters light with a fixed polarization.  
Therefore, one should pull the molecules out of the dark states to keep the optical cycling.
There are two ways -- one is remixing the magnetic substates using a magnetic field and the other is switching polarization of the light . 

Remixing the magnetic substates using a magnetic field relies on the magnetic precession of the molecules.
When a magnetic field is applied with an angle other than 0 or 90 degrees with respect to the quantum axis, the magnetic substates of the molecules precess around the magnetic field. 
The precession speed is determined by the strength of the magnetic field and the magnetic dipole moment of the molecule. 
For molecules with one valance electron with spin $S=1/2$, a magnetic field with less than 10 Gauss is enough to remix the states at the time scale of the optical cycling\cite{Shuman2009}.
The important thing here is that the magnetic field should not be parallel or perpendicular to the polarization of the laser light, which determines the quantum axis, to mix the magnetic substates. 

Switching the polarization of the light is more straight forward. 
Dark states for one polarization are bright states for the other polarization that is orthogonal to the initial polarization.
One needs to switch the polarization at the time scale of the excited state lifetime for the molecules to keep scattering photons without being trapped in dark states. 
Switching polarization of a laser at MHz rate can be easily achieved with a Pockel cell.

\begin{figure}[t!]
\includegraphics[width=17.0cm]{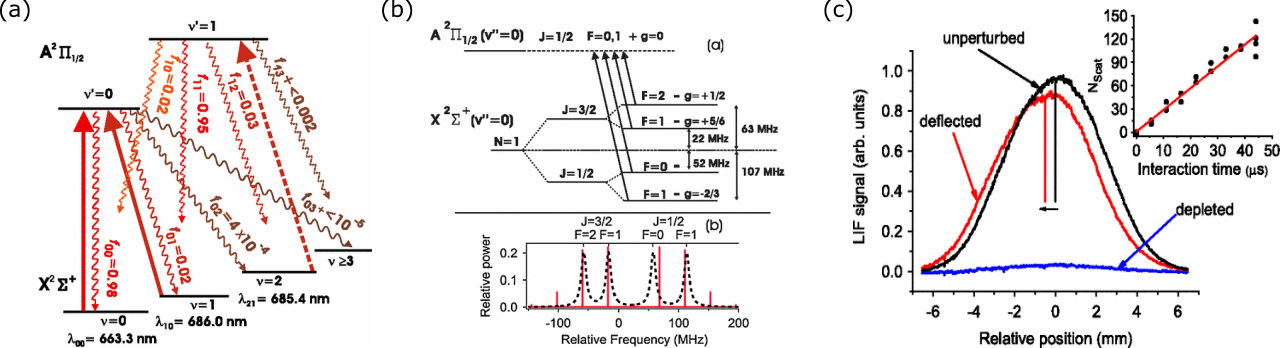}
\caption{(a) The vibrational energy structure of SrF molecules and the quasi-cycling transition. (b) Hyperfine structure of the $X(v=0, N=1)$ state of SrF molecules and the laser frequencies to address them (c) Deflection of SrF molecular beam due to photon cycling\cite{Shuman2009}.} 
\label{Fig:SrFEnergyStructure}
\end{figure}

In ref.\cite{Shuman2009}, authors demonstrated the first optical cycling of molecules and deflected a molecular beam of SrF molecules by photon kicks. 
Figure \ref{Fig:SrFEnergyStructure}(a) shows the electronic and vibrational energy structure of SrF molecules and laser frequencies relevant to optical cycling.
A laser at 663 nm (main laser) excites the molecules from the electronic and vibrational ground state ($X(v=0)$) to the electronic excited state with the same vibrational quantum number ($A(v'=0)$). 
As the molecules decay back to the electronic ground states, about 2\% of the molecules fall into the first excited vibrational state ($X(v=1)$).
Another laser at 686 nm (1$_\mathrm{st}$ repump laser) was employed to repump back the molecules in $X(v=1)$ state to $A(v'=0)$ state so that the molecules can scatter photons continuously.
With these two lasers, the molecules were able to cycling about 100 photons which is limited by the leakage to the second vibrational excited state in the electronic ground state. 
 
The main and the $1_\mathrm{st}$ repump laser should address all the states in the involved vibrational states.
As mentioned above, the rotational transition is closed by addressing a transition from $R=1$ state to $R=0$ state.
The good quantum numbers to express the rotational states for $X$ and $A$ state are $N$ and $J$ respectively, after combining the rotation with the electron's angular momentum ($X$ state) or with the electron's angular momentum and spin ($A$ state). 
Therefore, $X(N=1)$ to $A(J=1/2, +)$ transition forms a closed-cycling transition here, where $+$ indicates the parity of the $A$ state.
SrF molecule has an electron spin $S=1/2$ and a nuclear spin $I=1/2$.
The first excited rotational state $N=1$ in the $X$ state first couples to the electron spin to form $J=3/2$ and $J=1/2$ states. 
These two states again couple to the nuclear spin to result in four hyperfine states. 
 $A(J=1/2, +)$ state. 
The hyperfine states in the excited $A(J=1/2, +)$ state, in which the electron spin is already included, are not resolved due to the linewidth of the excited state.
Therefore, the main and the $1_\mathrm{st}$ repump lasers contained four frequencies generated by an electro-optic modulator (EOM) to address all the hyperfine states as shown in Figure \ref{Fig:SrFEnergyStructure}(b).

When the lasers described above were aligned perpendicular to the molecular beam towards one direction, the molecular beam was deflected by an amount proportional to the number of scattered photons (Figure \ref{Fig:SrFEnergyStructure}(c)). 
This was the first realization of optical cycle of molecules and showed the possibility for laser cooling molecules.


\section{Transverse cooling and two dimensional magneto-optical trap}
Laser cooling of molecules was first demonstrated along transverse direction of the molecular beam in Ref.\cite{Shuman2010}. 
Here, three lasers - the main, the 1st repump, and the 2nd repump lasers ($X(v=2)$ to $A(v'=1)$) - were employed to include $X(v=0,1,2)$ states and form a quasi-cycling transition. 
Theoretically, this enables upto about 2,500 photon scattering limited by the leakage to $X(v=3)$ state. 
The lasers were sent at an angle almost 90 degrees to the molecular beam and were about 75 times reflected back and force along 15 cm region to ensure enough interaction time between the laser and the molecules. 
The end of the cooling region, the lasers were retro-reflected to form a standing wave. 
A magnetic field was applied to remix the dark magnetic substates. 

\begin{figure}[t!]
\includegraphics[width=7.0cm]{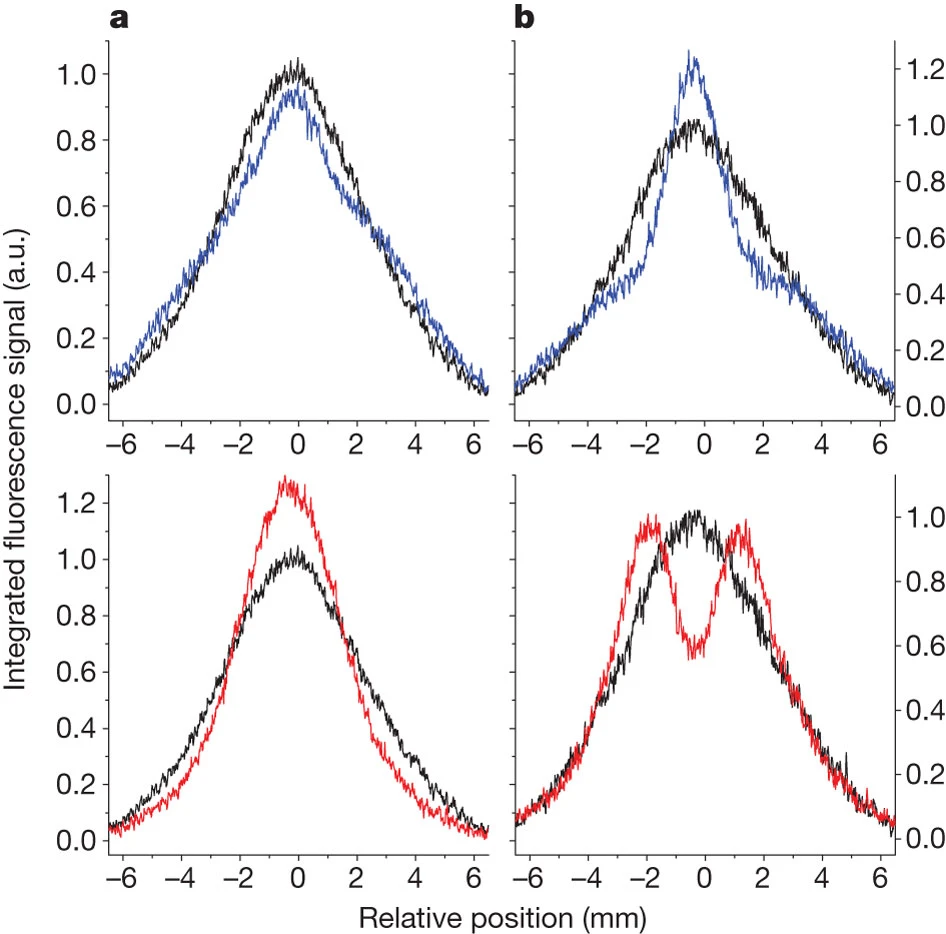}
\caption{Transverse cooling of SrF molecules\cite{Shuman2010}. Black lines show the transverse width of the unperturbed molecular beam. Blue(red) lines are the width of the molecular beam after interacting with blue-detuned(red-detuned) lasers. (a) When the magnetic field $B=5$ Gauss. Doppler cooling and heating were observed depending on the laser detuning. (b) When the magnetic field $B=0.6$ Gauss. Sisyphus cooling and heating were observed which had opposite dependence on the laser detuning compared to Doppler force.} 
\label{Fig:SrFTransverse}
\end{figure}

Figure \ref{Fig:SrFTransverse} summarizes the main result of the first transverse cooling of SrF molecules.
The black lines indicate the width of the unperturbed molecular beam 39 cm downstream from the molecular source. 
This measured width corresponds to the transverse velocity of the molecular beam because the divergence due to the transverse velocity is dominant over the initial width of the molecular beam at the source due to the long longitudinal distance that the molecules travel.
It was set to about 4 m/s by collimating apertures which corresponds to about 50 mK in temperature. 
For both Figure \ref{Fig:SrFTransverse}(a) and (b), red(blue) lines show the widths of the molecular beam after interacting with -1.5$\Gamma$(+1.5$\Gamma$)-detuned lasers where $\Gamma$ is the linewidth of the excited state. 
For $B=5$ Gauss, authors observed Doppler cooling and heating depending on the laser detuning (Figure. \ref{Fig:SrFTransverse}(a)). 
When the detuning was -1.5$\Gamma$, the width of the molecular beam decreased, indicating a significant cooling due to the Doppler force (red trace in Figure \ref{Fig:SrFTransverse}(a)).
The Doppler cooling force affected the entire width of the molecular beam because the Doppler force is effective over a large range of the velocity. 
The amount of cooling observed here corresponded to 500 to 1,000 photon scatters.
Doppler heating was observed as an increase of the molecules at outer locations with a blue-detuned laser of +1.5$\Gamma$.
When a smaller $B$ field of 0.6 Gauss was applied, Sisyphus cooling of molecules was achieved with the combination of the magnetic field and the standing wave of lasers at the end of the interaction region. 
The blue trace in Figure \ref{Fig:SrFTransverse}(b) shows a sharp peak in the center due to Sisyphus cooling force. 
Because the velocity range of the Sisyphus cooling force is much narrower than the maximum transverse velocity of the molecular beam, the Sisyphus cooling could only affect the central region of the molecular beam. 
When a red-detuned laser was applied under this weak $B$ field of 0.6 Gauss, Sisyphus heating was observed as accumulations of molecules at side positions where the Sisyphus force were no longer effective. 

\begin{figure}[t!]
\includegraphics[width=13.0cm]{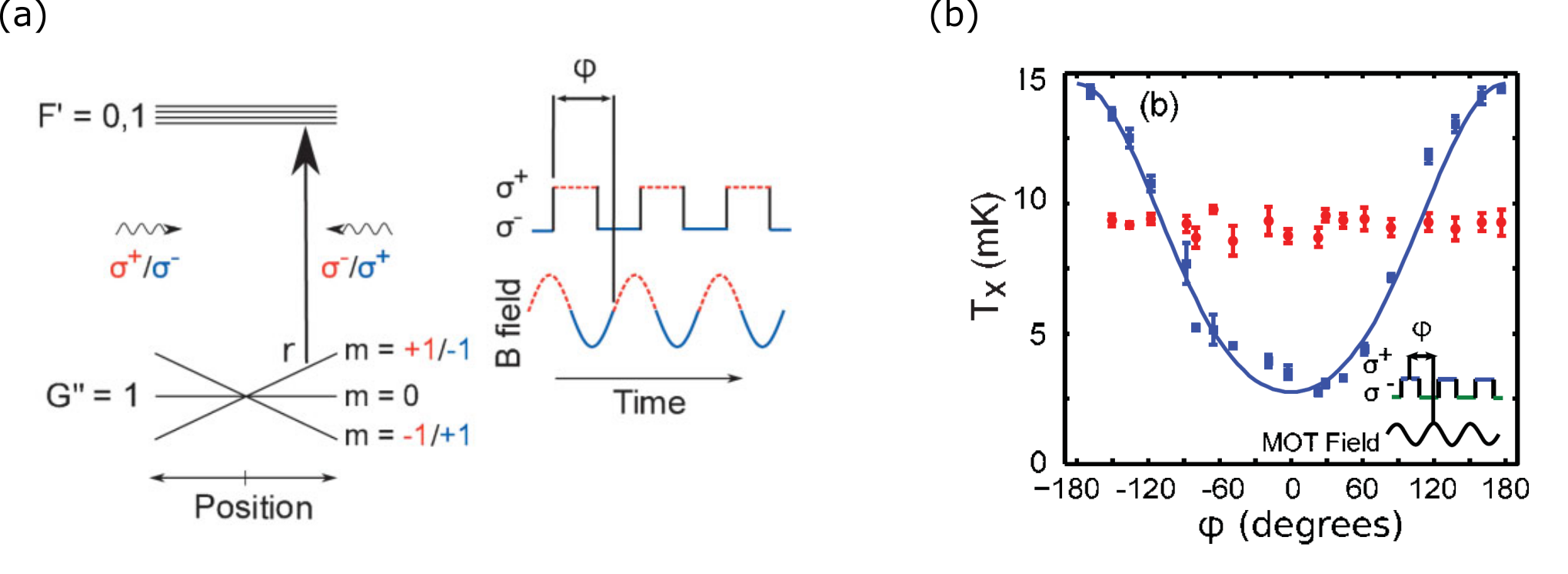}
\caption{(a) A schematic figure for a RF MOT. The polarizations of lasers and magnetic field should be switched at MHz time scale in order to scatter photons continuously and get trapping forces. (b) Transverse temperatures of the molecular beam after laser cooling. The transverse temperature of the unperturbed molecular beam was 25 mK set by collimating apertures. Red(blue) dots are for Doppler cooling(One dimensional MOT). Features of a MOT and an anti-MOT are clearly shown as a function of the relative phase between the magnetic field and the polarization of the lasers. } 
\label{Fig:YO2DMOT}
\end{figure}

After the realization of transverse laser cooling, two dimensional MOT was demonstrated with YO molecules\cite{Hummon2013}.
A MOT requires a magnetic field whose magnitudes are proportional to the distance from the origin and a pair of counter-propagating lasers with opposite circular polarizations to make space-dependent forces on the molecules.
The lasers are red-detuned so that they are only resonant with molecules with positive Zeeman shifts because the involved transition can be simplified as J=1 to J'=0 for a molecule due to its rotational structures.
The polarizations and the magnetic field are chosen so that the molecules scatter lasers that give momentum-kicks towards the origin, resulting in a restoring force.  
When one starts making a molecular MOT, the magnetic substates cause serious problems.
First, we cannot apply a remixing $B$ field because that will ruin the function of the MOT.
Therefore, to continue the photon cycling, we need to rapidly switch polarizations of the lasers at the time scale of MHz.
However, when the polarization is reversed, the laser now create anti-trapping forces. 
Therefore, the magnetic field should also be switched to the opposite directions at the same time scale of the polarization switching (RF MOT, Figure \ref{Fig:YO2DMOT}(a)).
The phase of the magnetic field and the polarization should match to generate trapping force. 

Authors of Ref.\cite{Hummon2013} demonstrated two dimensional magneto-optical force along the beam of YO molecules. 
The initial transverse temperature of the molecular beam was set to 25 mK by collimating apertures. 
As the transverse cooling experiment with SrF, three lasers were employed to form a quasi-cycling transitions. 
The lasers sent along two axes perpendicular to the molecular beam, interacting with the molecules for 275 $\mu$s. 
The detuning of the main laser was -5 MHz, which corresponds to about -$\Gamma$ of the excited state of YO.
The polarizations of the lasers and the direction of the $B$ field were switched at 2 MHz which matched to the optical pumping time scale.
Figure \ref{Fig:YO2DMOT}(b) shows the transverse temperature of the molecular beam after the molecules interact with the lasers along one axis.
Without the magnetic field, the molecules were Doppler cooled down to 10 mK (red data). 
Blue data show transverse temperatures after the molecules are cooled by the magneto-optical force.
The temperature varied from 3 mK to 15 mK depending on the relative phase between the switching of the laser polarizations and the magnetic field. 
One can clearly observe a MOT and an anti-MOT when the phase was 0 degrees and 180 degrees respectively.


\section{Longitudinal slowing}

To trap molecules in a MOT, the forward velocity of the molecular beam should be reduced down below the capture velocity of the MOT.
A typical slow molecular beam from a cyrogenic buffer-gas beam source has a forward velocity of 50 to a few hundred m/s depending on the mass of the species and the geometry of the buffer-gas cell\cite{Hutzler2012}.
However, the capture velocity of a molecular MOT ranges from 5 m/s upto 50 m/s depending on the species and one needs to bridge the gap between them by slowing the molecular beam. 
The most efficient way to do that is laser slowing where the molecules are slowed down by scattering a lot of counter-propagating photons.
Particularly, Zeeman slower is the most common method to slow down the atomic beam.
Zeeman slower uses a red-detuned single-frequency laser irradiated along the opposite direction to the atomic beam to slow down the atoms.
However, as the atoms slow down, the Doppler shift changes and the atoms cannot scatter the photons later on. 
Zeeman slower compensates the change of the Doppler shift by the Zeeman shift so that the atoms can scatter the photons continuously. 
Although there are recent research on building a Zeeman slower for molecules\cite{Kaebert2021}, it is not easy to make a Zeeman slower for molecules due to their complex energy structures.
Instead, two methods are mainly adopted to slow down the molecules using photon scattering -- white-light slowing and frequency chirping.

\begin{figure}[t!]
\includegraphics[width=10.0cm]{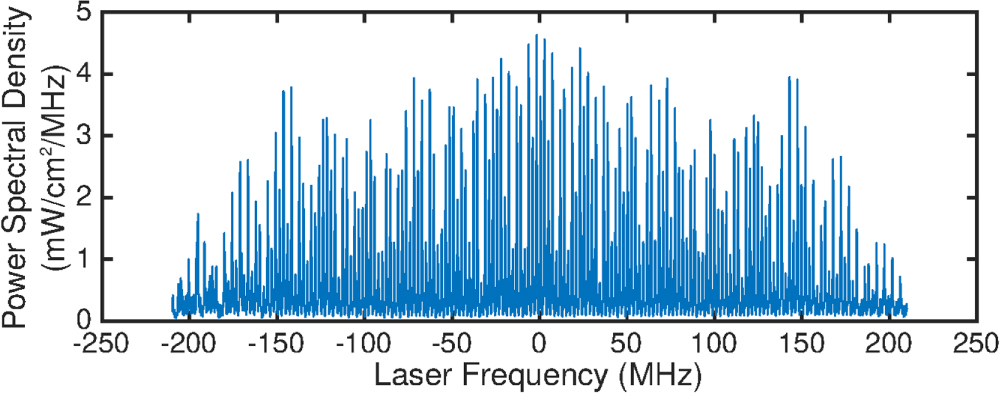}
\caption{An example of the frequency of the slowing laser for CaF after broadened by an overdriven EOM with a modulation index of about 17\cite{Hemmerling2016}. The resonant frequency of the EOM was set to 4.5 MHz, which is approximately half of the excited state linewidth of 8.29 MHz. The overall range of the frequency was about 400 MHz, which corresponds to the velocity range of 240 m/s for CaF.} 
\label{Fig:WhitelightEO}
\end{figure}

In White-light slowing, the frequency of the laser is broadened to cover a wide range of velocities of the molecules. 
Therefore, there always exist photons with resonant frequency to the molecules as they slow down and the radiation force can be provided continuously to the molecules. 
To slow the molecules from a few hundred m/s down to zero, the frequency of the laser should cover a range of over a few hundreds of MHz. 
An over-driven electro-optic modulator (EOM) is used to achieve this criteria (FIgure \ref{Fig:WhitelightEO}). 
The resonant frequency of the EOM is set to near the linewidth of the excited state so that molecules with any velocity can resonantly scatter photons. 

\begin{figure}[t!]
\includegraphics[width=7.0cm]{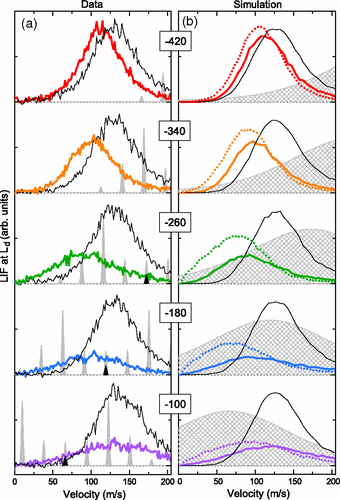}
\caption{Laser slowing of SrF molecules\cite{Barry2012}. Black lines indicate the initial forward velocity distribution of the molecular beam. The numbers written between (a) and (b) are the center detunings of the slowing lasers Grey peaks shown in (a) show the laser frequencies of the slowing laser converted to velocities of SrF. The grey area in (b) indicate the velocity-dependent force on molecules generated by the laser. After laser slowing, the distribution shifted towards slower velocities depending on the center frequency of the slowing lasers (colored lines in (a)). The measured velocity distributions agree well to the simulations that include the divergence of the molecular beam(colored lines in (b)). The dashed colored lines in (b) are the simulated results without the divergence of the molecular beam. } 
\label{Fig:SrFSlowing}
\end{figure}

Barry et al., demonstrated the first laser slowing of SrF molecules\cite{Barry2012}. 
As in the transverse laser cooling, they used three lasers for optical cycling and a magnetic field to remix the dark substates.
Figure \ref{Fig:SrFSlowing} shows how the velocity of the SrF molecules change for different center detunings of the main laser. 
The velocity distribution shifted towards a slower range after slowing (colored lines) compared to the initial distribution (black lines) and the average velocity shift increased as the detuning approached to the resonance.
When the detuning is close to the resonance, the amount of the detected molecules was reduced due to the divergence of the molecular beam, especially for slower molecules.  

\begin{figure}[t!]
\includegraphics[width=13.0cm]{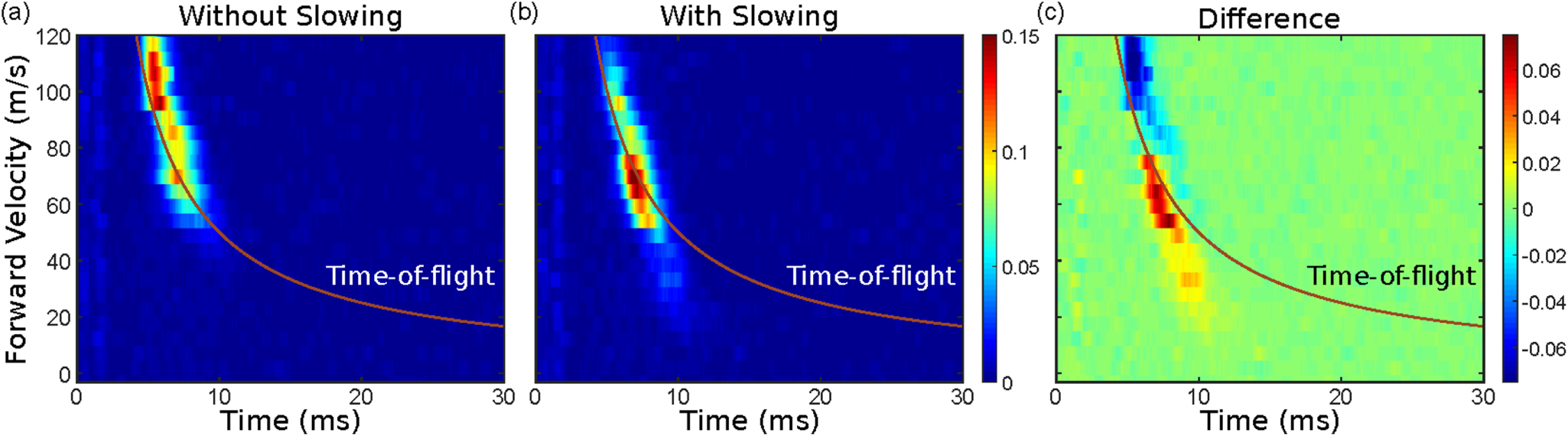}
\caption{Time-of-flight measurement of CaF molecules\cite{Hemmerling2016}. (a),(b) The velocities and arrival times of the initial molecular beam (a) and the laser-slowed molecular beam (b). The arrival times of the detected molecules agreed well with the expected time-of-flight line determined by the velocity and the traveling distance in (a). In (b), the overall velocities of the molecules shifted downwards compared to the initial distribution. Also, there are certain molecules who arrived earlier than the time-of-flight line, indicating they were slowed as they traveled. (c) the difference between (a) and (b). Blue(red) area shows the decrease(increase) of the molecular signal.} 
\label{Fig:CaFSlowing}
\end{figure}

Laser slowing could be confirmed more clearly when both the velocities and the arrival times of the molecules were measured. 
In Figure \ref{Fig:CaFSlowing}, the time-of-flight measurement was performed on CaF molecules before and after laser slowing\cite{Hemmerling2016}.
The laser-slowed molecules arrived earlier than the expected times determined by the velocity and the traveling distance, indicating that they first had higher velocities and slowed down to the detected velocity as they moved to the detection region. 
The difference between the initial beam and the laser-slowed beam indicated that molecules with velocities higher than 100 m/s were slowed down below 70 m/s, where the slowest molecules detected in this plot was 15 m/s. 

\begin{figure}[t!]
\includegraphics[width=10.0cm]{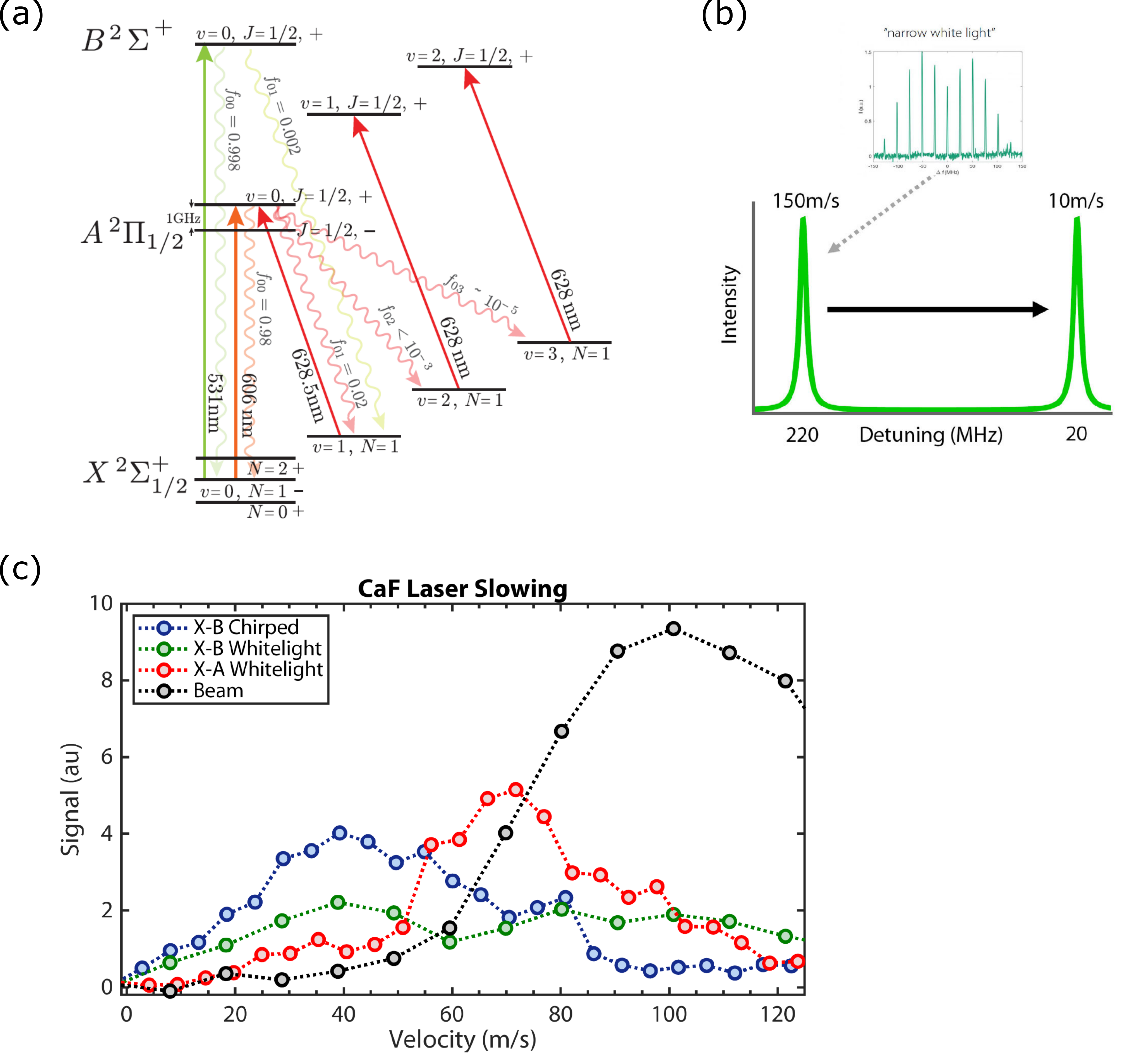}
\caption{(a) The relevant energy structure of CaF molecules and employed transitions for laser slowing\cite{Anderegg2017}. (b) A schematic figure for frequency chirping. The frequency of the slowing laser with sidebands for the hyperfine structure is rapidly changed for an amount of about 200 MHz to match the velocity change of the molecules as they slow down. (c) Velocity changes of CaF molecules after various laser slowing technique\cite{Anderegg2019a}} 
\label{Fig:CaFSlowingComparison}
\end{figure}

As people conducted experiments where molecules scattered a few thousands of photons, they realized that the photon scattering rate was greatly reduced because the main and the 1st repump laser shared the same excited states ($A$ state). 
The total number of states involved in the photon scattering through this excited state is $12(X(v=0)) + 12(X(v=1)) + 4(A(v'=0)) = 28$. 
In this case, the maximum scattering rate when all the transitions are saturated is $\Gamma/28$.
The scattering rate of the slowing laser could be increased if one can decouple the excited states of the main and the 1st repump lasers.
$B$ state of CaF, the second lowest electronic excited state, has even more diagonal FCFs compared to $A$ state (\ref{Fig:CaFSlowingComparison}(a)) which is good to be used as an alternative main slowing transition. 
The decay rate to $A$ state from $B$ state is negligible.
The efficiency of laser slowing was improved by adopting $X-B$ transition as the main and $X-A$ transition as the first repump transition as shown as green circles in Figure \ref{Fig:CaFSlowingComparison}(c).  

Although white-light slowing successfully slowed molecules down to near the capture velocity of the MOT, there are drawbacks. 
First, one needs a lot of laser power to get enough scattering rates because the laser power is distributed over a wide range of frequencies. 
Second, the deceleration happens regardless of the timing, position, and initial velocity of the molecular beam.
This results in a decreased number of molecules reaching to the region of interest for trapping molecules because molecules decelerated too early have more probability to diverge out or to turn around to the opposite direction.
To avoid these drawbacks, frequency chirping has been implemented\cite{Anderegg2019a}. 
In frequency chirping, the frequency of the slowing laser was rapidly changed to keep resonant to the molecular transition as the molecules slowed down (Figure \ref{Fig:CaFSlowingComparison}(b)). 
To cover the hyperfine structure, a few sidebands were added to the slowing laser. 
Typical chirping speed is about 200 MHz in a few ms, which is quite rapid and it can only be implemented to laser systems that allows a fast frequency modulation.
The efficiency of slowing increased a lot with the frequency chirping, resulting in more slow molecules that could be loaded into a MOT (blue circles in Figure \ref{Fig:CaFSlowingComparison}(c)).


\section{Three dimensional magneto-optical trapping}

\begin{figure}[t!]
\includegraphics[width=6.0cm]{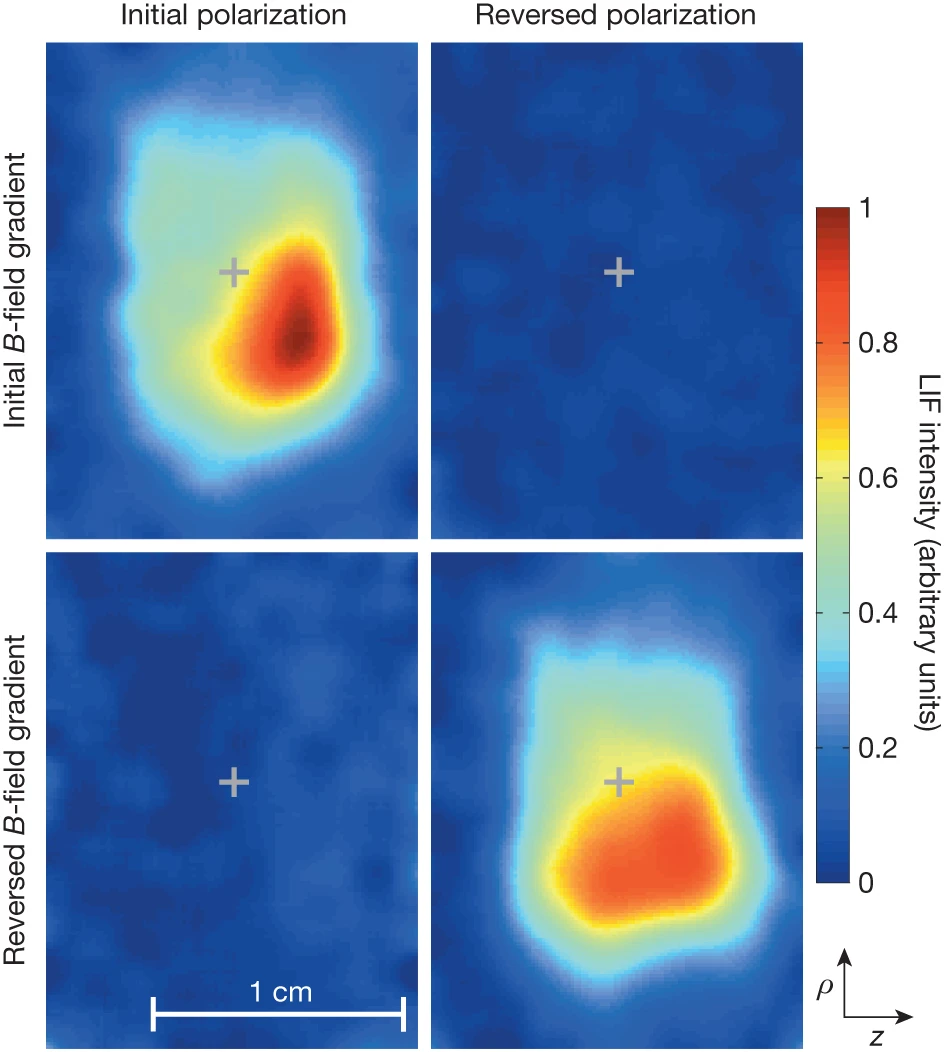}
\caption{The first DC MOT for SrF\cite{Barry2014}. The MOT only appeared when the direction of the $B$ field and the laser polarizations were matched to generate trapping forces.} 
\label{Fig:SrFMOT}
\end{figure}

To capture the molecules and cool them down to ultracold temperatures, a MOT is necessary.
As explained in Section III, molecules' magnetic substates in the ground state causes serious problems and a molecular MOT is impossible in a simple one dimensional picture unless using a RF MOT.
In three dimensional cases, however, the direction of the $B$ field and the laser polarization vary depending on the position. 
Therefore, as the molecules continuously move around, they see different $B$ field and laser polarizations and keep scattering photons, although at a slower rate.
As a result, the molecules can be trapped in a MOT even though the laser polarization and the $B$ field are not switched back and forth (DC MOT).
The first molecular MOT was achieved for SrF molecules in this way\cite{Barry2014}. 
As shown in Figure \ref{Fig:SrFMOT}, the MOT was big and fuzzy compared to a typical atomic MOT because of the reduced trapping force.
The number of trapped molecules was estimated to be 300 with the density of 600 cm$^{-3}$ and the temperature was 2 mK, which is a lot higher than the Doppler temperature of 160 $\mu$K. 

\begin{figure}[t!]
\includegraphics[width=15.0cm]{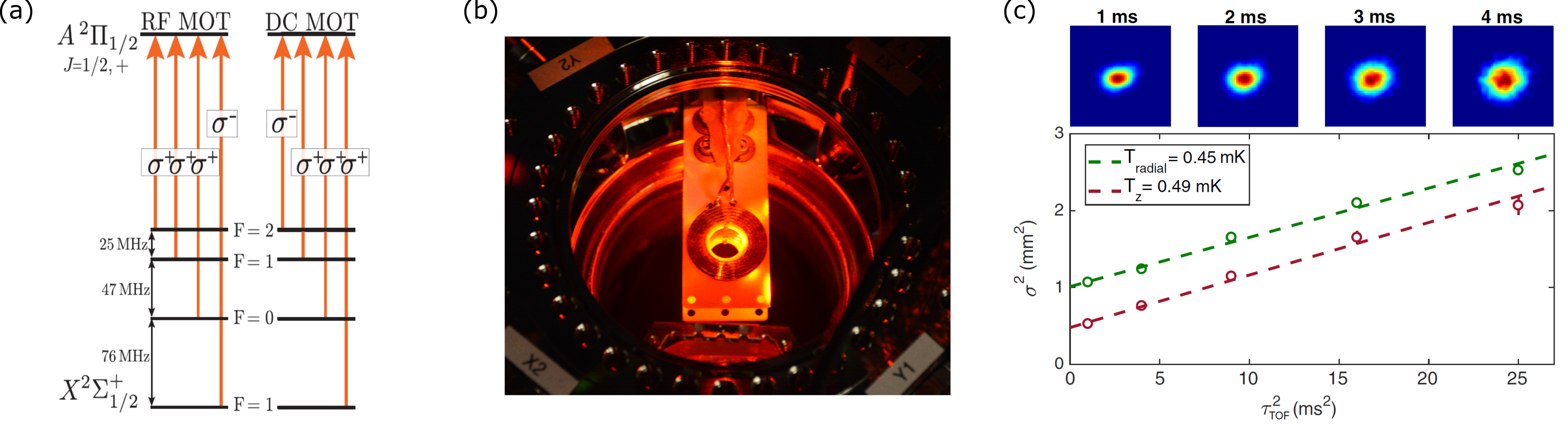}
\caption{(a) Optimum polarization configuration for a DC MOT and a RF MOT\cite{Anderegg2017}. RF MOT requires typical polarizations determined by the $g$-factors of the states. (b) A photo of the RF MOT coils inside the vacuum chamber. It was blackened after taking this photo to reduce the background photon scattering\cite{Chae2016}. (c) Time of flight measurement of a CaF RF MOT\cite{Anderegg2017}. In this paper, the lowest temperature of 340 $\mu$K was achieved, which is closed to the Doppler temperature of 200 $\mu$K.} 
\label{Fig:CaFMOT}
\end{figure}

A detailed simulation revealed that the trapping force is further reduced for molecules when the upper state has a much smaller $g$-factor than the ground state\cite{Tarbutt2015b, Tarbutt2015c}. 
Tarbutt et al. also found that a strong restoring force would be generated when the polarizations of the lasers were chosen differently as shown in Figure \ref{Fig:CaFMOT}(a) (DC MOT) taking into account optical pumping which causes the molecules to spend different times in different magnetic substates\cite{Tarbutt2015b}.
By following this strategy, the density of the SrF DC MOT was greatly increased to 4,000 cm$^{-3}$ with 500 molecules in the trap\cite{McCarron2015}.
Following SrF, CaF molecules were captured in a DC MOT using the optimized polarizations, resulting in $1.3 \times 10^4$ trapped molecules with a density of $1.6 \times 10^5$ cm$^{-3}$\cite{Truppe2017}. 
However, the temperatures were increased to 11 -- 13 mK in these DC MOTs with the optimized polarizations. 
This is due to a reduced damping coefficient of the MOT, resulted from polarization-gradient forces that oppose the Doppler force at low velocities.

Contrast to the DC MOTs, an RF MOT should be able to realize a high number and density of molecules as well as a low temperature because the RF MOT should have the same scattering rate and force to typical atomic MOTs. 
However, it is quite challenging to switch the current through typical anti-Helmholtz coils at MHz time scale to generate the desired $B$ field.
Typically, coils are installed outside of a vacuum chamber and therefore have sizes on the order of 10 cm with 10s of current going through them. 
These coils have large inductance and capacitance so that it is impossible to switching the current at MHz rate. 
Therefore, the coils should have a small size, and as a consequence, they should be installed inside the vacuum chamber and be UHV compatible.
Figure \ref{Fig:CaFMOT}(b) shows an example of the RF MOT coils installed in a vacuum chamber. 
The coil system was blackened after the photo was taken to reduce the background photon scattering.

The RF MOT for SrF molecules indeed had a lot better properties - it had 2,000 number of molecules with the density of $6 \times 10^4$ cm$^3$, and the temperature of 400 $\mu$K was achieved, which is close to the Doppler temperature\cite{Norrgard2016}. 
People also achieved a RF MOT for CaF molecules with the highest density of $7.3 \times 10^6$ cm$^{-3}$ and $1.03 \times 10^5$ molecules in the MOT. The lowest achieved temperature was 340 $\mu$K (Figure \ref{Fig:CaFMOT}(c)\cite{Anderegg2017}).

After the success of the MOTs for diatomic molecules with the simplest energy structures, people have succeeded to trap molecules with more complex energy structures in MOTs. 
YO molecules have an intermediate electronic state between the optical cycling transition. 
If molecules excited by the laser decay back to the electronic ground state through this intermediate state, they emit two photons and the molecules fall into different rotational states ($R=0,2$) with the opposite parity from the initial states. 
This stops the photon scattering.
Therefore, they should be repumped back to the initial states to keep the optical cycling.
Collopy et al. used microwaves that are resonant to $R=0$ to $R=1$ transition and $R=1$ to $R=2$ transition to directly pump back the loss to other rotational states\cite{Collopy2018}.
With these repumping microwaves, a RF MOT for YO molecules were achieved with $1.5 \times 10^4$ number of molecules in the trap at 4.1 mK.

The temperature of molecules in a MOT is several hundreds of $\mu$K, usually slightly higher than the Doppler temperature. 
At this temperature, thermal excitations of molecular internal states can be greatly ignored and molecules can be treated just as an atom using lasers and microwaves despite of its complexity. 


\section{Sub-Doppler cooling and transferring to conservative traps}

\begin{figure}[t!]
\includegraphics[width=15.0cm]{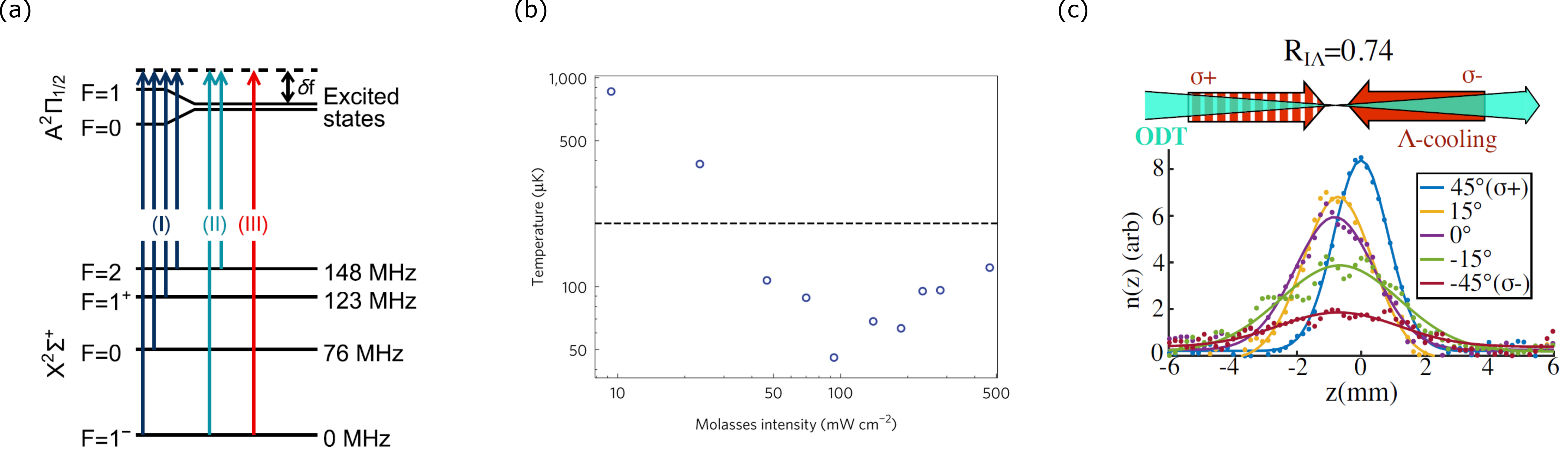}
\caption{(a) Various frequency configurations for sub-Doppler cooling\cite{Caldwell2019}. (b) Sub-Doppler cooling of CaF molecules using blue-detuned optical molasses\cite{Truppe2017}. (c) $\lambda$-enhanced cooling of SrF in an ODT\cite{Langin2021}. The temperature depended on the polarization of the ODT laser due to diffrential AC Stark shifts of the involved states.} 
\label{Fig:SubDopplerSchemes}
\end{figure}

To further harness and maximally use the quantum nature of the molecules, it is desired to cool down the molecules further.
Therefore, sub-Doppler cooling used for atoms can be immediately implemented to molecules.
Truppe et al., demonstrated the first sub-Doppler cooling by transferring the molecules into blue-detuned optical molasses ((I) in Figure \ref{Fig:SubDopplerSchemes}(a)) where molecules climb up an energy hill as they move from dark states to bright states (grey molasses)\cite{Truppe2017}.
The lowest temperature achieved here was 55 $\mu$K, which is about 1/4 of the Doppler temperature.

Even lower temperature was achieved using $\lambda$-enhanced grey molasses, in which the dark states are velocity-dependent ((II) in Figure \ref{Fig:SubDopplerSchemes}(a)).
A Temperature of 5 $\mu$K was achieved with CaF molecules using two blue-detuned lasers to form the $\lambda$-enhanced  cooling\cite{Cheuk2018}, as well as using only one blue-detuned laser\cite{Caldwell2019}.
$\lambda$-enhanced  cooling for CaF molecules was applied in optical dipole traps (ODTs) as well, achieving a higher temperatures of 20 $\mu$K compared to in free space due to the different AC Stark shifts for involved states\cite{Cheuk2018}.

In succession, $\lambda$-enhanced cooling of YO\cite{Ding2020} and SrF\cite{Langin2021} molecules were demonstrated, reaching 4 $\mu$K over a large range of $B$ field and 14 $\mu$K in an optical dipole trap with properly chosen polarizations respectively (Figure \ref{Fig:SubDopplerSchemes}(c)).

\begin{figure}[t!]
\includegraphics[width=12.0cm]{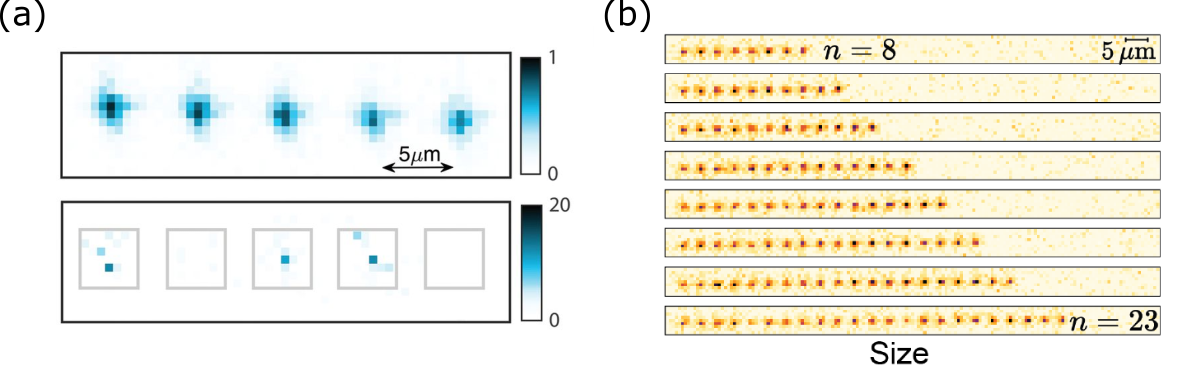}
\caption{An array of single CaF molecules in optical tweezers\cite{Anderegg2019b, Holland2022}. Rearranging the tweezer sites allowed to make a defect-free array of molecules.} 
\label{Fig:CaFTweezer}
\end{figure}

When molecules scatter photons in a MOT or in cooling procedure, their quantum states redistribute continuously within the cycling transitions. 
Therefore, transferring molecules into a conserved trap is necessary to manipulate single quantum state of molecules for many interesting applications. 
The most widely used traps are a magnetic trap and an ODT. 
The depth of the traps are about mK with a reasonable experimental setup that can be cooperative with a UHV chamber. 
With the addition of sub-Doppler cooling, molecules could be loaded into a conservative trap\cite{McCarron2018, Williams2018, Cheuk2018} as well as an optical lattice\cite{Wu2021} .
An array of single molecules trapped in optical tweezers and has been also achieved \cite{Holland2022, Anderegg2019b} (Figure \ref{Fig:CaFTweezer}), which allows deterministic quantum-state manipulation of single molecules.  

\section{Extension to polyatomic molecules}

The strategy for laser cooling of diatomic molecules can be readily extended to polyatomic molecules.
People have found most of the diatomic molecules with highly diagonal FCFs have their valence electron mostly orbiting around the metal atom like Ca and Sr (optical cycling center) in their electronic ground state and the lowest excited state. 
This indicates that even large molecules can be laser cooled if they have optical cycling centers -- consisted of one metal atom and something that is attached to it\cite{Augenbraun2020a, Dickerson2022, Lao2022, Yu2022}.
Because of the increased number of vibrational modes and the coupling between the vibration and the rotation, more lasers are required to form a quasi-cycling transition that was enough to capture the molecules. 
Although the scheme was technically complicated with many involved lasers, the basic idea is the same as the simple diatomic molecules. 

\begin{figure}[t!]
\includegraphics[width=14.0cm]{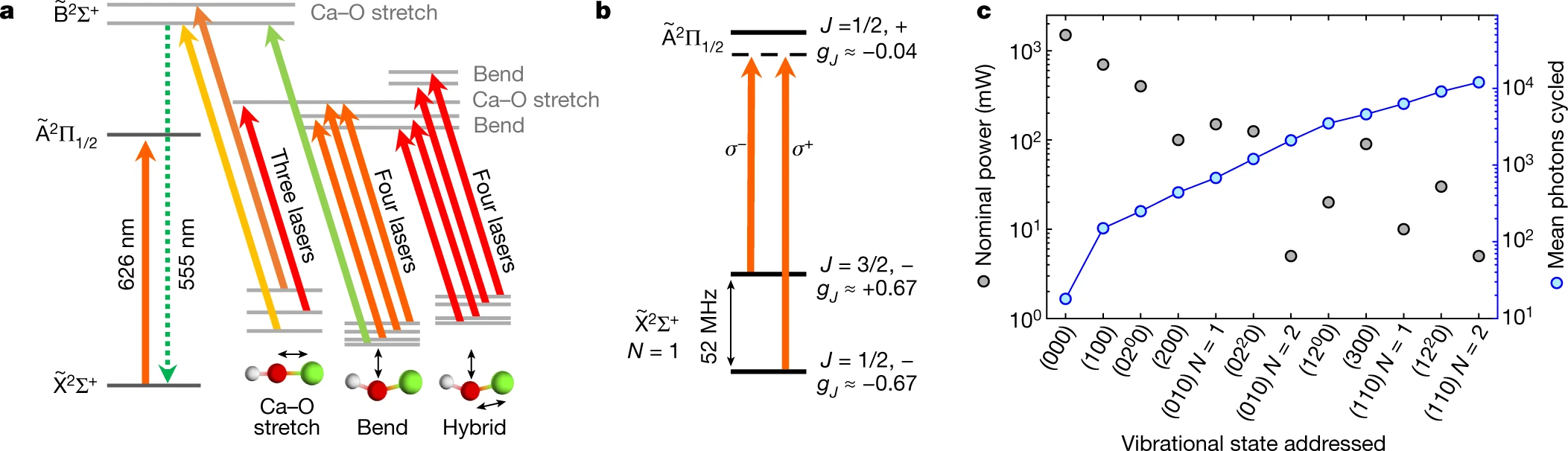}
\caption{Laser configuration for the MOT of CaOH molecules\cite{Vilas2022}. (a) The relevant energy structure of CaOH molecules and laser transitions employed to form a quasi-cycling transitions. (b) Polarizations of the MOT lasers. (c) The required power of the employed lasers to saturate the transitions (black) and the mean number of cycled photons when all preceding repump lasers were added from left to right (blue)}
\label{Fig:CaOHCooling}
\end{figure}

Experimentally, one-dimensional cooling of YbOH\cite{Augenbraun2020}, CaOCH$_3$\cite{Mitra2020}, and a MOT and sub-Doppler cooling of CaOH\cite{Vilas2022} have recently been achieved.
12 lasers were implemented for the CaOH MOT to address all the vibrational and bending modes to cycle upto $10^4$ photons as shown in Figure \ref{Fig:CaOHCooling}. 

\section{CONCLUSIONS}

The overall principles, techniques, and development of laser cooling of molecules are reviewed. 
After achieving the quasi-cycling transitions with multiple lasers, which has been greatly eased with the development of laser technology over the last decades, molecules can be cooled and manipulated just as atoms.
However, the number of trapped molecules are still orders of magnitudes smaller than atomic species.
This is mainly due to the poor production of molecules at the beginning and less efficiency of slowing stage.
Many efforts are on-going to trap more molecule as well as broaden trapped molecular species.
With the increasing community, laser cooled molecules will enable diverse research in quantum simulation/computation, quantum chemistry, as well as precision measurement to test fundamental rules that govern our universe. 


\begin{acknowledgments}

This work is supported by National Science Foundation of Korea (Grant No. 2020R1A4A1018015, 2021R1C1C1009450, 2021M3H3A1085299, 2022M3E4A1077340, 2022M3C1C8097622). 

\end{acknowledgments}


\bibliographystyle{unsrt}
\bibliography{References-LaserCoolingReview}

\end{document}